\RequirePackage{lineno}
\documentclass[aps,prc,twocolumn, floatfix]{revtex4}
\usepackage{graphicx}
\usepackage{dcolumn}
\usepackage{bm}
\usepackage{array}
\usepackage{multirow}
\usepackage[colorlinks=true, linkcolor=blue, citecolor=blue, urlcolor=blue]{hyperref}

\begin{document}


\title{Probing $\Lambda$ potential at high and low densities via \(^3_{\Lambda}\)H production in C+C reactions}

\author{Gao-Chan Yong$^{1,2,3}$}
\email[]{yonggaochan@impcas.ac.cn}
\author{J. L. Rodr\'{\i}guez-S\'{a}nchez$^{4,5}$}
\author{Yapeng Zhang$^{1,2,3}$}
\affiliation{
$^1$Institute of Modern Physics, Chinese Academy of Sciences, Lanzhou 730000, China\\
$^2$School of Nuclear Science and Technology, University of Chinese Academy of Sciences, Beijing 100049, China\\
$^3$State Key Laboratory of Heavy Ion Science and Technology, Institute of Modern Physics, Chinese Academy of Sciences, Lanzhou 730000, China\\
$^4$CITENI, Campus Industrial de Ferrol, Universidade da Coru\~{n}a, E-15403 Ferrol, Spain\\
$^5$IGFAE, Universidad de Santiago de Compostela, E-15782 Santiago de Compostela, Spain
}

\begin{abstract}
The production of Lambda hyperons and light hypernuclei in heavy-ion collisions provides critical insights into the nuclear equation of state (EoS) and hyperon interactions in dense matter, addressing the longstanding ``hyperon puzzle'' in neutron star physics. Using the INCL+ABLA and AMPT-HC models, we systematically investigate C+C collisions at beam energies of 1.1 and 1.9 GeV/nucleon to explore the sensitivity of hyperon and hypernuclei yields to the EoS and hyperon potential. Our results reveal that Lambda hyperon production is predominantly influenced by the nuclear EoS, while light hypernuclei formation (e.g., \(^3_{\Lambda}\)H) exhibits stronger sensitivity to the hyperon potential at high/low densities with the incident beam energies of 1.1/1.9 GeV. The study provides a framework for future experiments at facilities like HIAF and FAIR/GSI to resolve the hyperon puzzle, thus advancing our understanding of quantum chromodynamics in high-density regimes.

\end{abstract}

\maketitle

\section{Introduction}
The study of hyperons (baryons with strangeness content) in extreme nuclear environments, such as neutron stars and relativistic heavy-ion collisions, provides critical insights into the fundamental interactions in dense matter and the equation of state (EoS) of nuclear matter under extreme conditions. Hyperons, particularly $\Lambda$, $\Sigma$, and $\Xi$ particles, are predicted to emerge in neutron star interiors at densities of 2 $\sim$ 3 times nuclear saturation density due to the high chemical potentials of nucleons, significantly softening the EoS and reducing the maximum neutron star mass \cite{1}. However, the recent discovery of two solar masses' pulsars (e.g., PSR J0348+0432 \cite{2}) challenges this paradigm, creating the so-called ``hyperon puzzle'': the apparent contradiction between the theoretical prediction of hyperon-induced EoS softening and the observational evidence of massive neutron stars \cite{3}. This puzzle underscores the urgent need to refine our understanding of hyperon-nucleon (YN) and hyperon-hyperon (YY) interactions, which remain poorly constrained due to limited experimental data on hypernuclei and hyperon investigations \cite{4}.

Relativistic heavy-ion collisions (HICs) in terrestrial laboratories serve as a tool to probe the properties of dense matter and the in-medium interactions of hyperons. The production of hypernuclei and hyperons in HICs offers a unique opportunity to extract information about the hyperon potentials in nuclear matter, which directly influence the EoS. For instance, the binding energies of single- and double-$\Lambda$ hypernuclei, such as $^{6}_{\Lambda\Lambda}$He \cite{5}, provide critical constraints on the $\Lambda-\Lambda$ interaction and the coupling strengths of hyperons to mesonic fields in relativistic mean field (RMF) models \cite{1, 6}. These constraints are vital for constructing unified EoS models that reconcile hyperon presence with the two solar masses' neutron star observations \cite{1}. Furthermore, the analysis of hyperon yields and flow patterns in HICs can shed light on the density dependence of hyperon potentials, which may be also sensitive to EoS and isospin asymmetry of nuclear matter \cite{7}.

Despite advances in phenomenological and microscopic approaches --- such as RMF models \cite{1}, chiral effective field theory \cite{8}, and Brueckner-Hartree-Fock calculations \cite{9} --- significant uncertainties persist in the YN and YY interactions. For example, the $\Lambda$ potential in symmetric nuclear matter, is well constrained by hypernuclear data, but the $\Lambda$ potential in pure neutron matter, varies widely depending on the assumed vector meson couplings \cite{1}. Additionally, the repulsive $\Sigma$-nucleus potential inferred from the absence of $\Sigma$-hypernuclei \cite{4} and the poorly understood $\Xi$-nucleus interaction \cite{10} further complicate the EoS modeling. These uncertainties propagate into predictions for neutron star properties, such as radii and cooling behavior, and the viability of exotic phases like quark matter or kaon condensates \cite{11}.

This study aims to bridge the gap between hypernuclear experiments, heavy-ion collision observables, and neutron star astrophysics by systematically investigating the relationship between hyperon production in HICs, hyperon potentials, and the nuclear EoS. By calibrating YN/YY interactions to hypernuclear data and HIC results, we seek to develop a consistent EoS framework capable of explaining both terrestrial experiments and astrophysical observations. Such efforts will not only address the hyperon puzzle but also enhance our understanding of QCD in the high-density regime, paving the way for future experiments at facilities like FAIR/GSI, J-PARC and HIAF in China \cite{12}. In this study, we have conducted C+C reactions at beam energies of 1.1 and 1.9 GeV per nucleon, as this reaction will be carried out at both HIAF and FAIR/GSI facilities \cite{122}. The study reveals that C+C reactions near the Lambda hyperon production threshold (the laboratory beam energy required for nucleon-nucleon collisions to produce a Lambda hyperon is approximately 1.58 GeV) can generate not only hyperons but also light hypernuclei. Both hyperon production and light hypernuclei formation are influenced to varying degrees by the EoS and hyperon potential.

\section{Model Descriptions}
%
The Li\`{e}ge intranuclear-cascade (INCL) model is a sophisticated framework designed to simulate collisions between projectiles (e.g., nucleons, pions, hyperons, and light ions) and target nuclei \cite{incl2002, incl13, incl14}. It integrates classical physics principles with quantum-mechanical features such as Fermi motion, realistic spatial density distributions, and Pauli blocking to account for initial collision dynamics. The model employs an energy- and isospin-dependent nucleon potential and a constant isospin-dependent hyperon potential, both characterized by Woods-Saxon density profiles. Recent advancements in the INCL++ (v6.33) version include the capability to simulate hyper-remnant formation, tracking baryon number, charge, energy, momentum, angular momentum, and strangeness conservation \cite{incl22strange}. The INCL model treats collisions as a sequence of relativistic binary hadron-hadron interactions, dynamically updating particle positions and momenta over time. It also incorporates pion and strangeness production mechanisms, making it suitable for studying hyperon-related phenomena in nuclear matter \cite{incl22strange20}.

Coupled with the INCL model, the ABLA07 code describes the de-excitation processes of excited remnants formed during collisions \cite{abla07}. ABLA07 simulates the emission of $\gamma$-rays, neutrons, light charged particles, intermediate-mass fragments (IMFs), and fission products. Originally developed in FORTRAN, it has been adapted to C++ for compatibility with the GEANT4 framework. Recent extensions enable the simulation of hyperon emissions and cold hypernuclei formation \cite{incl22strange}. The combined INCL+ABLA framework provides a comprehensive approach to investigating hyperon-nucleon interactions and hyperon dynamics in nuclear medium.

On the other hand, the AMPT-HC model, an enhanced version of the multi-phase transport (AMPT) model \cite{AMPT2005}, incorporates hadronic mean-field potentials and conducts pure hadron cascade simulations \cite{cas2021,yongrcas2022,yongplb2023,phase2024}. It integrates initial density and momentum distributions of nucleons within colliding nuclei, determined by Skyrme-Hartree-Fock calculations using Skyrme M$^{\ast}$ force parameters, and the Fermi momentum derived via the local Thomas-Fermi approximation. The model reflects recent experimental insights into the nucleon momentum distribution, including a high-momentum tail reaching up to approximately twice the local Fermi momentum. Potentials for nucleons, resonances, hyperons, and their antiparticles are implemented using the test-particle method.
A Skyrme-like baryon mean field, or referred to as the EoS, was employed, as detailed in Ref.~\cite{cas2021}. For strange baryons's mean fields, the quark counting rule is applied, suggesting interactions through their non-strange components. Free elastic cross sections for proton-proton ($\sigma_{pp}$), neutron-proton ($\sigma_{np}$), and neutron-neutron ($\sigma_{nn}$) are determined by experimental data, with $\sigma_{nn}$ assumed equivalent to $\sigma_{pp}$ at similar center of mass energy. All other baryon-baryon free elastic cross sections are presumed equivalent to the nucleon-nucleon elastic cross section at the same center of mass energy. An experimental energy-dependent nucleon-nucleon inelastic total cross-section is utilized at lower energies \cite{phase2024}. In this study, the production of light hypernuclei is investigated using a phase-space coalescence approach \cite{gibuu08,yong2009}, with the fundamental assumption that $\Lambda$ hyperons and neutrons share equivalent coalescence dynamics.

In the context of hypernuclei formation, other hybrid dynamical-statistical approaches (e.g., SMM) combining transport models (e.g., UrQMD, DCM) with statistical decay frameworks have been extensively studied \cite{smm2024}. Additionally, advanced fragment recognition algorithms like FRIGA, which utilize simulated annealing to identify isotopes and hypernuclei in transport simulations, also provide critical insights into coalescence dynamics and hyperon potential dependencies \cite{friga2019}.
Compared to other models, the INCL+ABLA model employed in this study incorporates more detailed calculations related to thermal residual excitation energy, such as quantum tunneling effects and experimental separation energies for particle emission, along with additional quantum effects. Meanwhile, the ABLA de-excitation model places greater emphasis on dynamic fission processes and quantum statistical effects in particle emission. As for the AMPT-HC model, it focuses on finer details of isospin-dependent nucleon initialization and single-particle mean fields, including isospin-dependent nucleon initialization, isospin-dependent nucleon mean fields, isospin-dependent scattering cross-sections, and other related aspects.

\section{Results and Discussions}
\begin{figure}[t]
\centering
\includegraphics[width=0.45\textwidth]{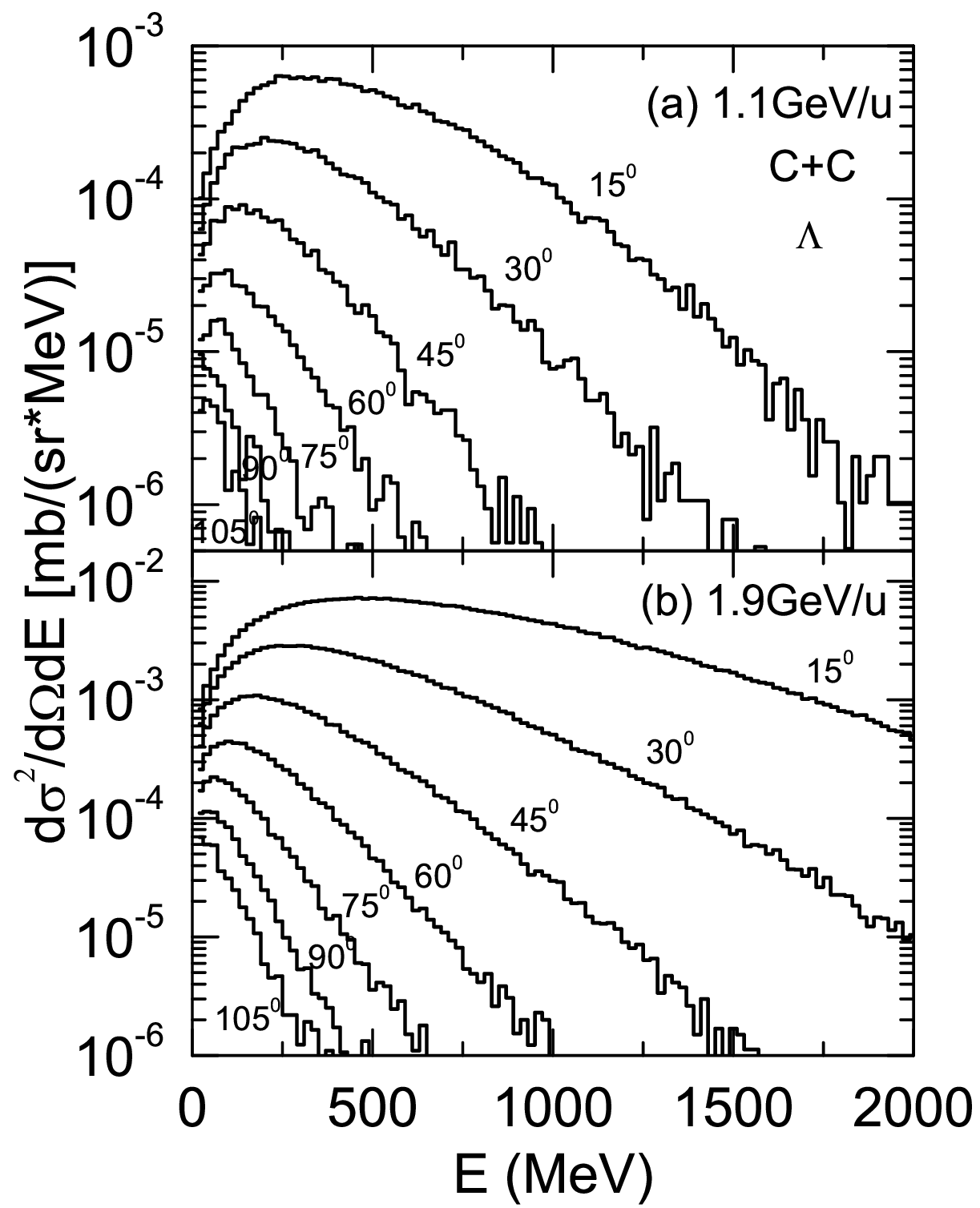}
\vspace{0.25cm}
\caption{Double differential cross-section of Lambda hyperon production as a function of kinetic energy in the $^{12}$C+$^{12}$C reaction, with beam energies of (a) 1.1 GeV/nucleon and (b) 1.9 GeV/nucleon in the laboratory frame. Simulated by the INCL+ABLA.} \label{doubleLam}
\end{figure}
To study the production of Lambda hyperons in heavy-ion collisions, it is useful to provide the distribution of Lambda hyperon production with respect to emission angle and kinetic energy for detection purposes. Figure~\ref{doubleLam} shows the double differential cross-section of Lambda hyperon production (without those originating from $\Sigma^{0}$ hyperon decay) as a function of hyperon kinetic energy for C+C collisions at beam energies of 1.1 and 1.9 GeV per nucleon. It is evident that as the polar angle increases, the distribution of Lambda hyperons decreases; moreover, as the kinetic energy increases, the number of Lambda hyperons also rapidly decreases. From the upper panel of Figure~\ref{doubleLam}, it can be seen that despite the beam energy being 1.1 GeV, the produced Lambda hyperons can reach up to 2 GeV, especially at small polar angles. Comparing the upper and lower panels of Figure~\ref{doubleLam}, it is clear that the number of above-threshold Lambda hyperons produced at 1.9 GeV beam energy is significantly higher than that at the below-threshold 1.1 GeV beam energy; the production of Lambda hyperons at higher beam energies is more inclined towards forward angles. Note that here, the hyperons produced through heavy-ion collisions, as well as the subsequently mentioned hypernuclei, may be harnessed as secondary beams to perform hyperon/hypernucleus-nucleus scattering experiments \cite{Simone23}. Such experiments hold significant importance for constraining the hyperon potential in a model-independent manner \cite{hyperpot2024}.

%
\begin{figure}[t]
\centering
\vspace{0.25cm}
\includegraphics[width=0.45\textwidth]{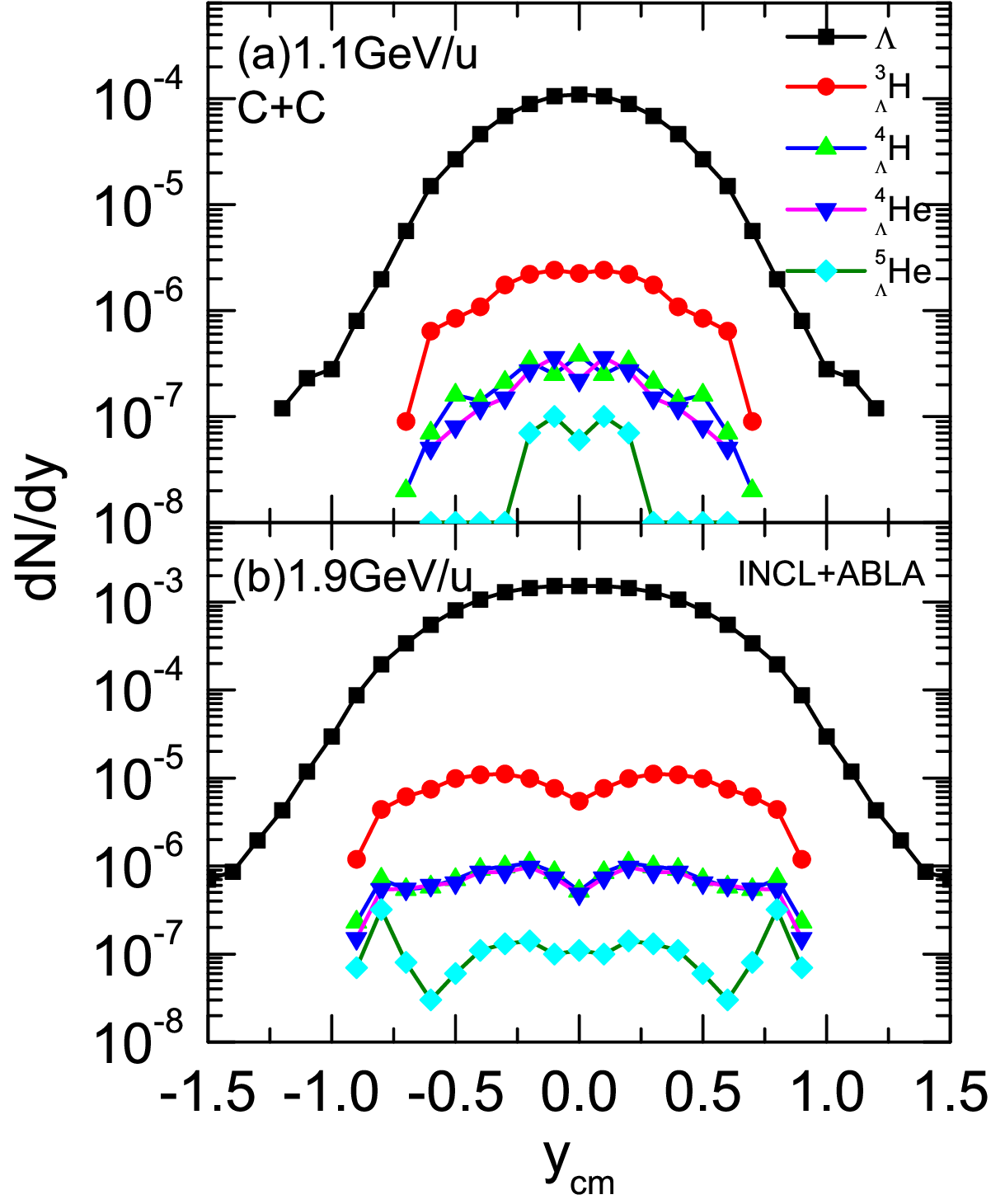}
\vspace{0.25cm}
\caption{Rapidity distribution (in the center-of-mass frame) of Lambda and Lambda hypernuclei in the $^{12}$C+$^{12}$C reaction, with beam energies of (a) 1.1 GeV/nucleon and (b) 1.9 GeV/nucleon in the laboratory frame. Here, the values of positive rapidity are a reflection of the negative rapidity values. Simulated by the INCL+ABLA.} \label{dNdy}
\end{figure}
To conveniently investigate the physics related to Lambda hyperon production in heavy-ion collisions, Figure~\ref{dNdy} presents the rapidity distributions of Lambda hyperons and light hypernuclei production. Here, the reason we use the INCL+ABLA model to predict the production of light hypernuclei is primarily because, at such low beam energies and within such a small reaction system, the yield of light hypernuclei is extremely low. The AMPT-HC model requires a vast number of events, whereas the INCL+ABLA model can conveniently perform these calculations. It can be seen that as the mass number of the particles increases, their yields decrease evidently. Overall, the number of Lambda hyperons and light hypernuclei produced in C+C collisions at the above-threshold beam energy of 1.9 GeV is significantly higher than that at the below-threshold energy of 1.1 GeV. Additionally, it can be observed that for C+C collisions at higher collision energies, the production of heavier light hypernuclei mainly occurs in the large rapidity region. This is the region of the ``spectator'' distribution in heavy-ion collisions, where heavier light hypernuclei are primarily produced.

\begin{figure}[t]
\centering
\includegraphics[width=0.45\textwidth]{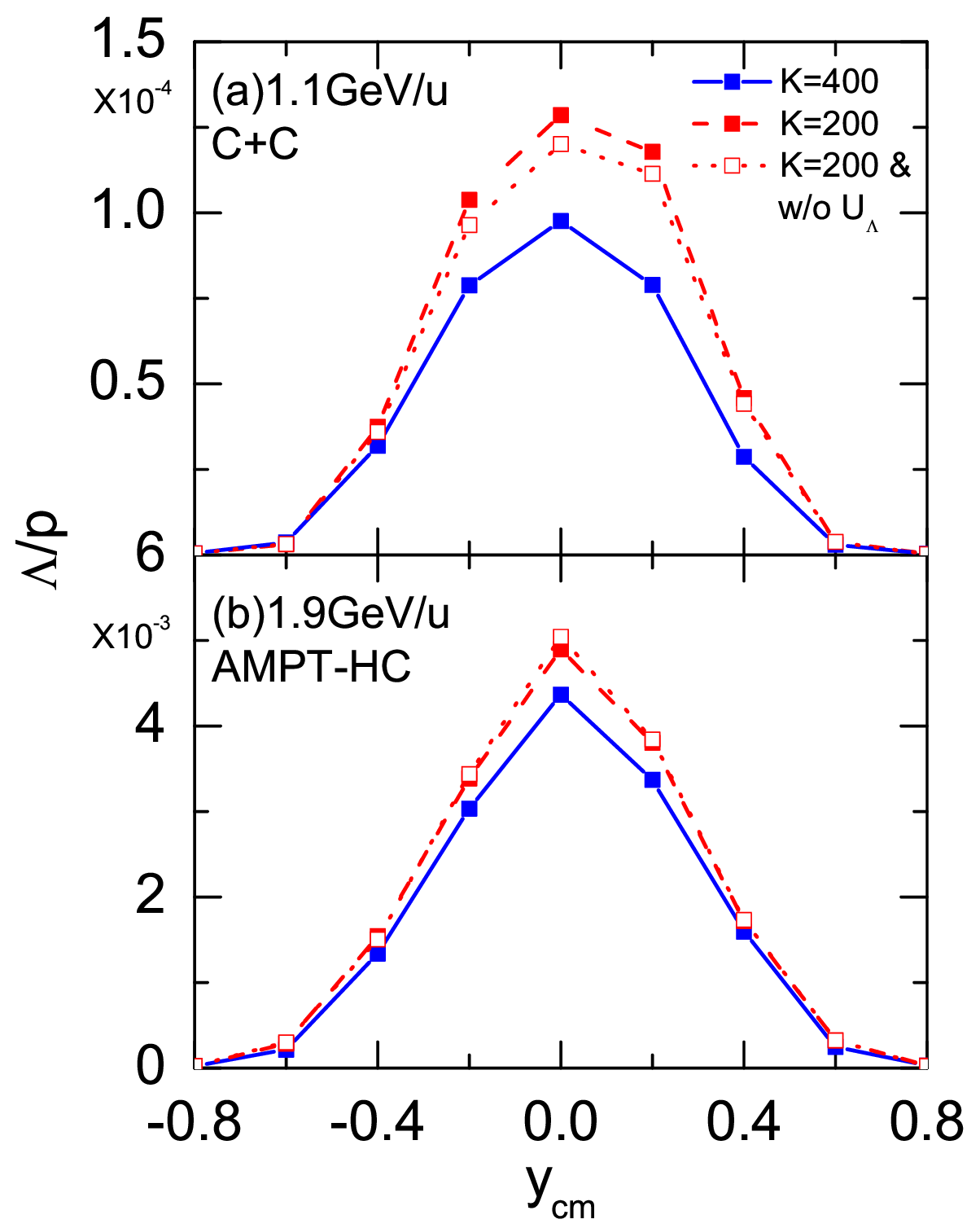}
\vspace{0.25cm}
\caption{Lambda hyperon to proton ratio as a function of rapidity (in the center-of-mass frame). Different lines represent different EoSs (i.e, the incompressibility coefficient K= 200, 400 MeV) and hyperon potentials in the $^{12}$C+$^{12}$C reaction, with beam energies of (a) 1.1 GeV/nucleon and (b) 1.9 GeV/nucleon in the laboratory frame. Simulated by the AMPT-HC.} \label{ratio}
\end{figure}
The study of the EoS and hyperon potential has always been one of the forefront topics in the field of heavy-ion collisions. Is the production of hyperons in heavy-ion collisions more sensitive to the nuclear matter EoS or the hyperon potential? To explore these questions, we first present Figure~\ref{ratio}, which shows the yield ratio of Lambda hyperons to protons as a function of rapidity. The advantage of ratio observables over single-variable observables lies in their ability to reduce theoretical uncertainties and systematic errors. It is evident that the yield ratio of Lambda hyperons to protons is more sensitive to the nuclear EoS than to the hyperon potential. This is mainly because Lambda hyperons are primarily produced during the maximum compression phase of nuclear matter, a process that significantly depends on the stiffness of the nuclear EoS. The hyperon potential only plays a role during the expansion phase of nuclear matter, whereas the nuclear EoS is influential throughout the entire heavy-ion collision process, both during compression and expansion phases. Therefore, pre-equilibrium emitted Lambda hyperons are more sensitive to the nuclear EoS than to the hyperon potential. Furthermore, from Figure~\ref{ratio}, it can also be seen that the yield ratio of Lambda hyperons to protons for below-threshold Lambda hyperon production is more sensitive to the nuclear EoS. The fact that Lambda hyperons produced in heavy-ion collisions are more sensitive to the EoS than to the hyperon potential is understandable, as Lambda hyperons are, after all, secondary particles and should be more sensitive to the EoS experienced by their parent particles.

\begin{table}[t!]
\vspace{0.25cm}
\resizebox{0.475\textwidth}{!}{
\begin{tabular}{|c|c|c|c|c|}
    \hline
    EoS and $U_{\Lambda}$ &  \multicolumn{2}{c|}{1.1 GeV} &  \multicolumn{2}{c|}{1.9 GeV}  \\ \hline\hline
    soft+$U_{\Lambda}$ & 0.34$\times10^{-5}$ & 0 & 0.17$\times10^{-4}$ & 0  \\ \hline
    stiff+$U_{\Lambda}$ & 0.17$\times10^{-5}$ & 50\% & 0.16$\times10^{-4}$ & 6\%  \\ \hline
    soft w/o $U_{\Lambda}$ & 0.16$\times10^{-5}$ & 53\% & 0.88$\times10^{-5}$ & 48\%  \\ \hline
    soft with $U^{\rho>\rho_{0}}_{\Lambda}$ & 0.22$\times10^{-5}$ & 35\% & 0.18$\times10^{-4}$ & 6\%  \\ \hline\hline  
    soft+$U_{\Lambda}$ & 0.215$\times10^{-3}$ & 0 & 0.533$\times10^{-2}$ & 0  \\ \hline
    stiff+$U_{\Lambda}$ & 0.166$\times10^{-3}$ & 23\% & 0.475$\times10^{-2}$ & 11\%  \\ \hline
    soft w/o $U_{\Lambda}$ & 0.238$\times10^{-3}$ & 10\% & 0.58$\times10^{-2}$ & 8\%  \\ \hline
    soft with $U^{\rho>\rho_{0}}_{\Lambda}$ & 0.208$\times10^{-3}$ & 3\% & 0.538$\times10^{-2}$ & 1\%  \\ \hline
\end{tabular}}
\caption{Number of $^{3}_{\Lambda}$H and $\Lambda$ production in C+C reaction. The first 4 rows represent the production of $^{3}_{\Lambda}$H, while the last 4 rows represent the production of $\Lambda$. The ``soft'' indicates a EoS with incompressibility coefficient K= 200 MeV while ``stiff'' indicates K= 400 MeV. $U^{\rho>\rho_{0}}_{\Lambda}$ denotes the Lambda potential at only high densities. The percentages indicate the degree of change relative to the case K= 200 MeV. Given by the AMPT-HC plus coalescence afterburner.}
\label{yields}
\end{table}

To further explore the effects of the hyperon potential, it is now necessary to investigate whether the production of light hypernuclei is sensitive to the hyperon potential; and for the production of light hypernuclei, whether the high-density hyperon potential plays a dominant role or the low-density hyperon potential does. The reason for conducting this separate investigation is that a significant body of research has shown that the production of clusters in heavy-ion collisions is noticeably influenced by the density of the surrounding medium --- for example, a high medium density tends to inhibit cluster production \cite{Ono2019}. Therefore, the hyperon potential in the low-density region may play a more significant role in the production of light hypernuclei in heavy-ion collisions.

Considering the sharp decline in the yield of light hypernuclei with increasing mass number, and to obtain better statistical significance, only $^{3}_{\Lambda}$H is studied here. Table~\ref{yields} lists the sensitivities of the production numbers of $^{3}_{\Lambda}$H in C+C reactions under the two beam energy conditions to the EoS and the hyperon potential. For comparison, the production of $\Lambda$ is also listed below. From the numbers in the table, it can be seen that for the production of $^{3}_{\Lambda}$H, under the 1.1 GeV condition, the hyperon potential effect is slightly greater than the EoS effect; the high-density hyperon potential effect is about twice that of the low-density hyperon potential effect; under the 1.9 GeV condition, the hyperon potential effect is also significantly greater than the EoS effect; but the high-density hyperon potential effect is significantly lower than the low-density hyperon potential effect. For hyperon production, under the 1.1 GeV condition, the EoS effect is roughly twice that of the hyperon potential effect; the low-density hyperon potential effect is about twice that of the high-density hyperon potential effect; under 1.9 GeV, the EoS effect is also slightly greater than the hyperon potential effect; the high-density hyperon potential effect is significantly lower than the low-density hyperon potential effect.

In short, for the production of light hypernuclei, the hyperon potential effect is greater than the EoS effect, but whether it is dominated by high-density or low-density hyperon potential depends on the beam energy. For hyperon production, the EoS effect is greater than the hyperon potential effect, with the low-density hyperon potential effect being dominant. The possible reason is that hyperons are mainly produced during the compression stage of the C+C collision, thus primarily influenced by the EoS, while the hyperon potential effect is mainly manifested during the later expansion process, thus having a smaller effect; whereas for the production of light hypernuclei, according to the coalescence model, light hypernuclei are mainly produced during the later expansion process, therefore the EoS effect is less significant than the hyperon potential effect. As for why the production of $^{3}_{\Lambda}$H is sensitive to the high-density hyperon potential at a beam energy of 1.1 GeV but shifts sensitivity to the low-density hyperon potential at 1.9 GeV, this is because, as shown in Figure~\ref{dNdy}, $^{3}_{\Lambda}$H's are predominantly produced in the mid-rapidity region (corresponding to the high-density regime) at 1.1 GeV, while their production transitions to the non-mid-rapidity region (associated with the low-density regime) at 1.9 GeV.

Here, we emphasize that the current study on hyperon and light hypernuclei production in the light reaction system \(^{12}\text{C}+^{12}\text{C}\) near the hyperon production threshold energy lacks existing experimental data or theoretical calculations for direct comparison, as this is the first investigation into hyperon and light hypernuclei production in low-energy, light systems and their sensitivity to the nuclear EoS and high-/low-density hyperon potentials. We hope that our current results will soon be compared with relevant experiments at HIAF and FAIR/GSI or with analogous theoretical studies. Notably, one of our key conclusions --- that hyperon production is predominantly sensitive to the nuclear EoS rather than the hyperon potential --- aligns with our prior research \cite{cas2021}. Meanwhile, the finding that light hypernuclei (e.g., \(^{3}_{\Lambda}\text{H}\)) exhibit sensitivity to high-density hyperon potentials at sub-threshold beam energies and to low-density potentials at above-threshold energies represents a novel contribution of this work. We anticipate that similar conclusions will emerge from studies based on other theoretical models in the near future.

Of course, the above conclusions may be model-dependent. For instance, using different equations of state, different forms of the hyperon potential, varying models for the production of light hypernuclei, employing heavier reaction systems, or using higher or lower collision energies could all potentially impact the results of the study. Therefore, the purpose of this research is to urge colleagues in this field to conduct more comprehensive and systematic studies on the use of hyperon or light hypernuclei to investigate and constrain the stiffness of the nuclear matter equation of state and the hyperon potential.

The next phase of our work will focus on analyzing experimental data released by the HypHI collaboration at GSI Laboratory, monitoring future FAIR-related experiments, tracking new developments at the WASA@FRS facility, and preparing for the upcoming Super-FRS apparatus. Notably, hypernucleus production research has been incorporated into the STAR experiment program at Brookhaven National Laboratory and multiple experiments at CERN's Large Hadron Collider (LHC). These initiatives will enable us to extract hyperon potential strengths under both low- and high-density conditions through experimental validation in the near future, marking a critical step toward resolving the neutron star hyperon puzzle.

\section{Conclusions}

This study elucidates the interplay between hyperon production, hypernuclei formation, and the nuclear equation of state in the upcoming C+C collisions. Key findings include:
%
The \(\Lambda\)/proton yield ratio is more sensitive to the nuclear EoS than to the hyperon potential, as hyperon production occurs predominantly during the high-density compression phase. A softer EoS (K=200 MeV) significantly enhances \(\Lambda\) yields at sub-threshold energies (1.1 GeV/nucleon), whereas its effect becomes less pronounced at higher energies (1.9 GeV/nucleon).
%
Light hypernuclei production, exemplified by \(^3_{\Lambda}\)H, is strongly influenced by the hyperon potential during the expansion phase and the effect of the hyperon potential exceeds that of the EoS, with high-density potentials dominating at 1.1 GeV/nucleon and low-density potentials at 1.9 GeV/nucleon.
%
%
These results bridge hypernuclear physics, heavy-ion collision experiments, and neutron star astrophysics, offering a pathway to resolve the hyperon puzzle.

\section{Acknowledgments}
This work is partly supported by the National Natural Science Foundation of China under Grant Nos. 12275322, 12335008, 12475133 and CAS Project for Young Scientists in Basic Research YSBR-088. J.L.R.-S. is thankful for the support provided by the Regional Government of Galicia under the program ``Proyectos de excelencia'' Grant No. ED431F-2023/43, and by the ``Ram\'{o}n y Cajal'' program under Grant No. RYC2021-031989-I funded by MCIN/AEI/10.13039/501100011033 and ``European Union NextGenerationEU/PRTR''.

\end{document}